\documentclass[runningheads]{llncs}
\usepackage[T1]{fontenc}
\usepackage{graphicx}
\usepackage{booktabs}
\usepackage{bm}
\usepackage{amstext}
\usepackage{subfigure}
\usepackage{mathtools}
\usepackage{amsmath}
\usepackage{enumitem}

\usepackage{amssymb}
\usepackage{multirow}
\usepackage{tcolorbox}
\usepackage{colortbl}
\usepackage{wrapfig}
\usepackage[misc]{ifsym}
\usepackage[colorlinks,linkcolor=blue]{hyperref}
\begin{document}

\title{Learning Social Graph for Inactive User Recommendation}

\author{Nian Liu\inst{1} \and
Shen Fan\inst{1} \and
Ting Bai \inst{2} \and
Peng Wang \inst{1} \and
Mingwei Sun \inst{1} \and
Yanhu Mo  \inst{1} \and
Xiaoxiao Xu \inst{1} \and
Hong Liu \inst{1} \and
Chuan Shi \inst{2} \Letter
}
\authorrunning{N. Liu et al.}
%
\institute{Alibaba Group Holding Limited, China \and Peng Cheng Laboratory, China \\
\email{\{nianliu1998, baiting0317, chuanshi1978\}@gmail.com \{fanshen.fs, paulwong.wp, sunmingwei.smw, moyanhu.myh, xiaoxiao.xuxx, liuhong.liu\}@alibaba-inc.com}
}
\maketitle              
\pagestyle{empty}
\begin{abstract}

Social relations have been widely incorporated into recommender systems to alleviate data sparsity problem. However, raw social relations don't always benefit recommendation due to their inferior quality and insufficient quantity, especially for inactive users, whose interacted items are limited. In this paper, we propose a novel social recommendation method called LSIR (\textbf{L}earning \textbf{S}ocial Graph for \textbf{I}nactive User \textbf{R}ecommendation) that learns an optimal social graph structure for social recommendation, especially for inactive users. LSIR recursively aggregates user and item embeddings to collaboratively encode item and user features. Then, graph structure learning (GSL) is employed to refine the raw user-user social graph, by removing noisy edges and adding new edges based on the enhanced embeddings. Meanwhile, mimic learning is implemented to guide active users in mimicking inactive users during model training, which improves the construction of new edges for inactive users. Extensive experiments on real-world datasets demonstrate that LSIR achieves significant improvements of up to 129.58\% on NDCG in inactive user recommendation. Our code is available at~\url{https://github.com/liun-online/LSIR}.

\end{abstract}
\section{Introduction}
With the rise of social media, there's growing interest in social recommendation~\cite{graphrec}, which uses social relations to improve recommendation systems~\cite{diffnet}. 
Compared to general recommender systems~\cite{ngcf}, social recommendation helps mitigate data sparsity issue, especially in cold-start setting~\cite{DBLP:InfNet,DBLP:conf/cikm/LiuSCLWZWZGZ21}. Inactive users\footnote{In a broad sense, `cold-start users'~\cite{gope2017survey} are a subset of `inactive users'. Inactive users can include both those who have been on the platform for a long time but have limited behaviors in the past, as well as newly registered users.} interact little with items on transaction platforms, making it challenging for precise recommendation. 

Most existing social recommendation models operate solely on the raw social graph~\cite{graphrec,diffnet,mhcn}, with little effort to examine the quality of social relations on recommendation performance. \textit{Are raw social relations adequately effective for inactive user recommendation?} Utilizing real social relations from one of China's largest e-commerce platforms, we analyze their impact on user behavior, particularly for inactive users. These analyses, uncover two flaws of raw social relations: (1) \textbf{Inferior quality}. The quality of raw social relations is generally inferior in real industrial scenario. Poor-quality social relations don't guarantee beneficial social information intake for users. (2) \textbf{Insufficient quantity}. Inactive users generally have fewer social relations than active users. This means less social information is available to inactive users. Combined with limited user-item interactions, inactive users receive very little information from both user and item aspects.



Above all, we realize that raw social relations are far from optimal for inactive users, requiring refinement for improved social recommendation. However, efficiently and effectively refining the social graph is technically challenging. In the Graph Neural Network domain~\cite{DBLP:journals/tnn/WuPCLZY21}, graph structure learning (GSL) is commonly utilized to learn an optimal graph structure~\cite{lds,sublime}. During standard GSL procedure, models are trained to filter noisy edges between existing neighbors and establish superior connections with long-range nodes. Existing GSL approaches~\cite{prognn} generally optimize the possibilities between every two nodes, incurring costly computational complexity. In our scenario, we could employ GSL on the raw social graph, deleting useless neighbors and adding sufficient and high-quality users for inactive users. However, we should constrain the search scope for potential structures due to scalability concerns.

In this paper, we explore the effect of social relations on inactive user recommendation, and incorporate GSL into social recommendation. We propose a novel model to \textbf{\textit{L}}earn \textbf{\textit{S}}ocial graph for \textbf{\textit{I}}nactive users \textbf{\textit{R}}ecommendation, named \textbf{\textit{LSIR}}. Specifically, users and items firstly recursively aggregate embeddings from each other, encoding collaborative signals. Using these aggregated embeddings, we refine the raw user-user social graph through delicately devised U2U deletion and U2C addition. U2U deletion retains high-quality existing social neighbors by removing noisy connections, while U2C addition adds connections between users and active interest clusters.  These enhance inactive user representations with remaining purified relations and new connections to active interest clusters. Meanwhile, only generating new connections from representative active users avoids high computing complexity compared to traditional GSL settings requiring probability computation between every two nodes. Finally, mimic learning manipulates active users to `mimic' the distribution of inactive users by maximizing similarity between pseudo inactive users and corresponding active interest clusters. Thus, the proposed LSIR can get more precise training on connecting inactive users and active users. Our contributions are summarized as follows:
\begin{itemize}[leftmargin=*]
    \item To the best of our knowledge, this is the first attempt to study the impact of social relations for social recommendation. We observe that inactive users tend to have limited social connections, some of which are low-quality, adversely affecting recommendation performance. 
    \item We propose LSIR, incorporating graph structure learning into social recommendation to address this issue. By refining the raw user-user social graph, LSIR enhances social recommendation through noisy connection deletion and beneficial connection addition. 
    \item We validate the effectiveness and scalability of LSIR against state-of-the-art methods on public and industrial datasets. LSIR demostrates substantial improvements up to 129.58\% on NDCG for inactive user recommendation.
\end{itemize}

\begin{figure*}[t]
\centering
\subfigure[Quality Analysis]
{
 	\begin{minipage}[b]{.28\linewidth}
        \centering
        \includegraphics[width=\linewidth,height=0.7\linewidth]{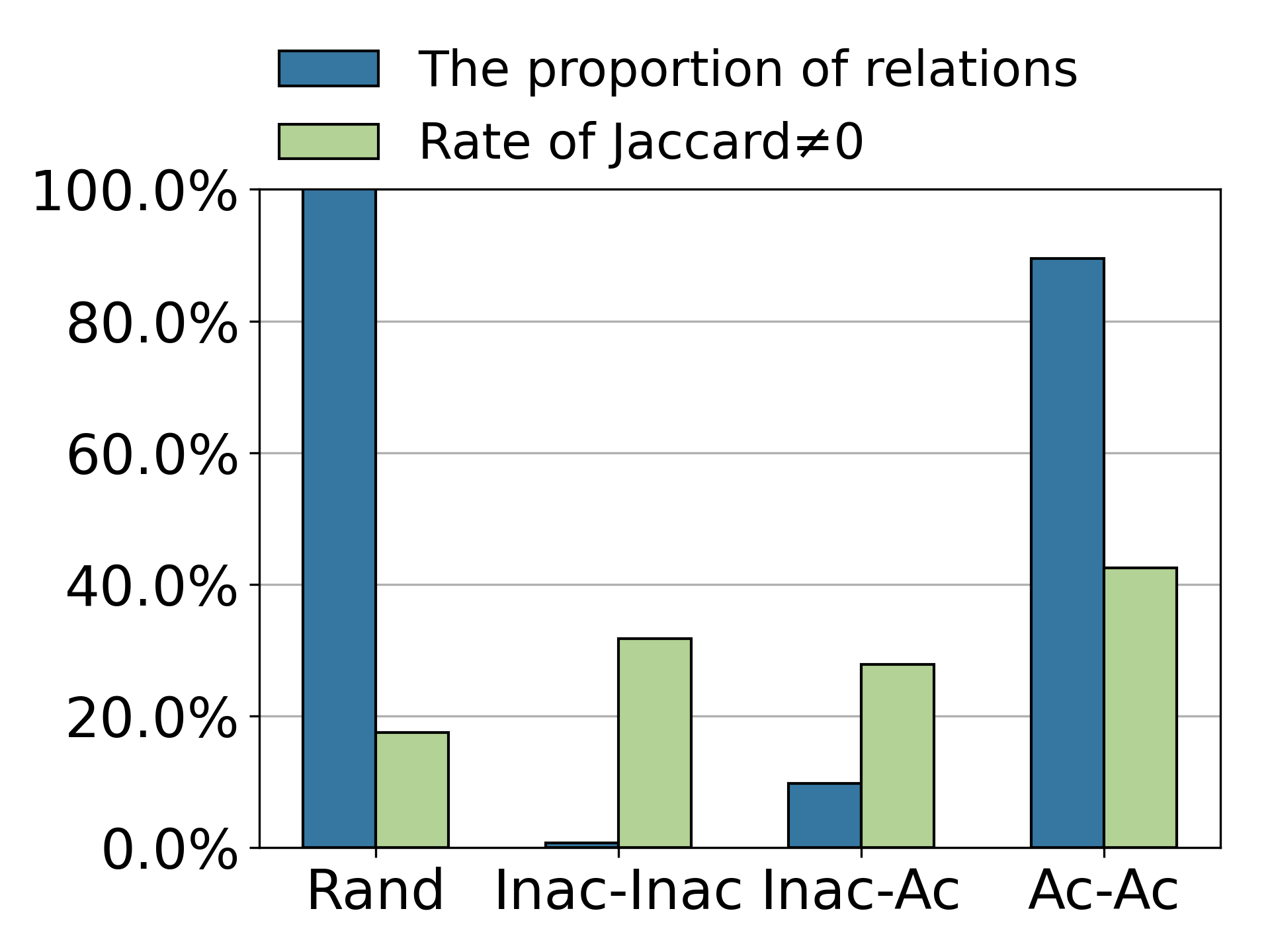}
    \end{minipage}
}
\subfigure[Quantity Analysis]
{
 	\begin{minipage}[b]{.28\linewidth}
        \centering
        \includegraphics[width=1\linewidth,height=0.7\linewidth]{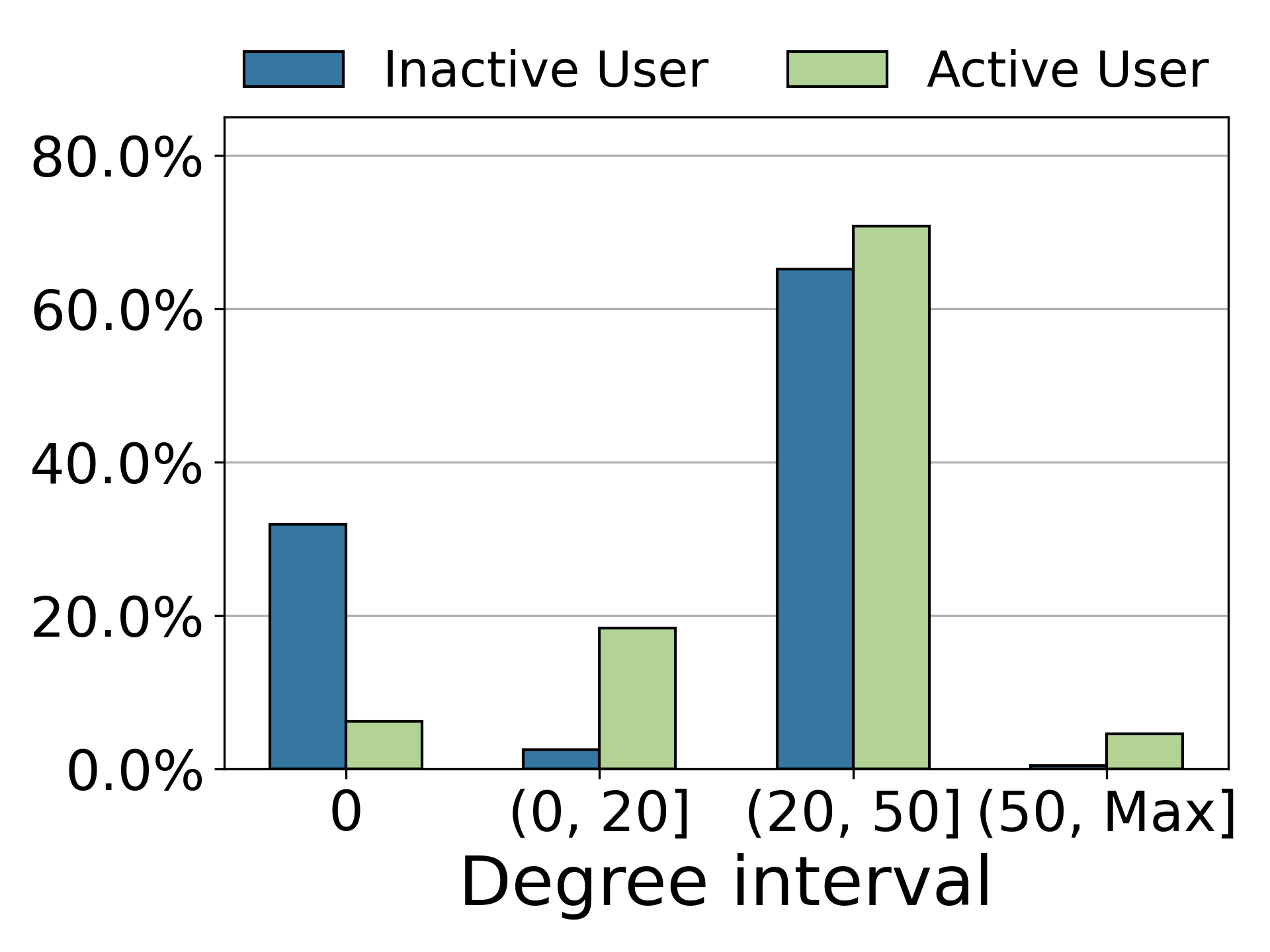}
    \end{minipage}
}
\subfigure[New Relation Effect]
{
 	\begin{minipage}[b]{.28\linewidth}
        \centering
        \includegraphics[width=1\linewidth,height=0.7\linewidth]{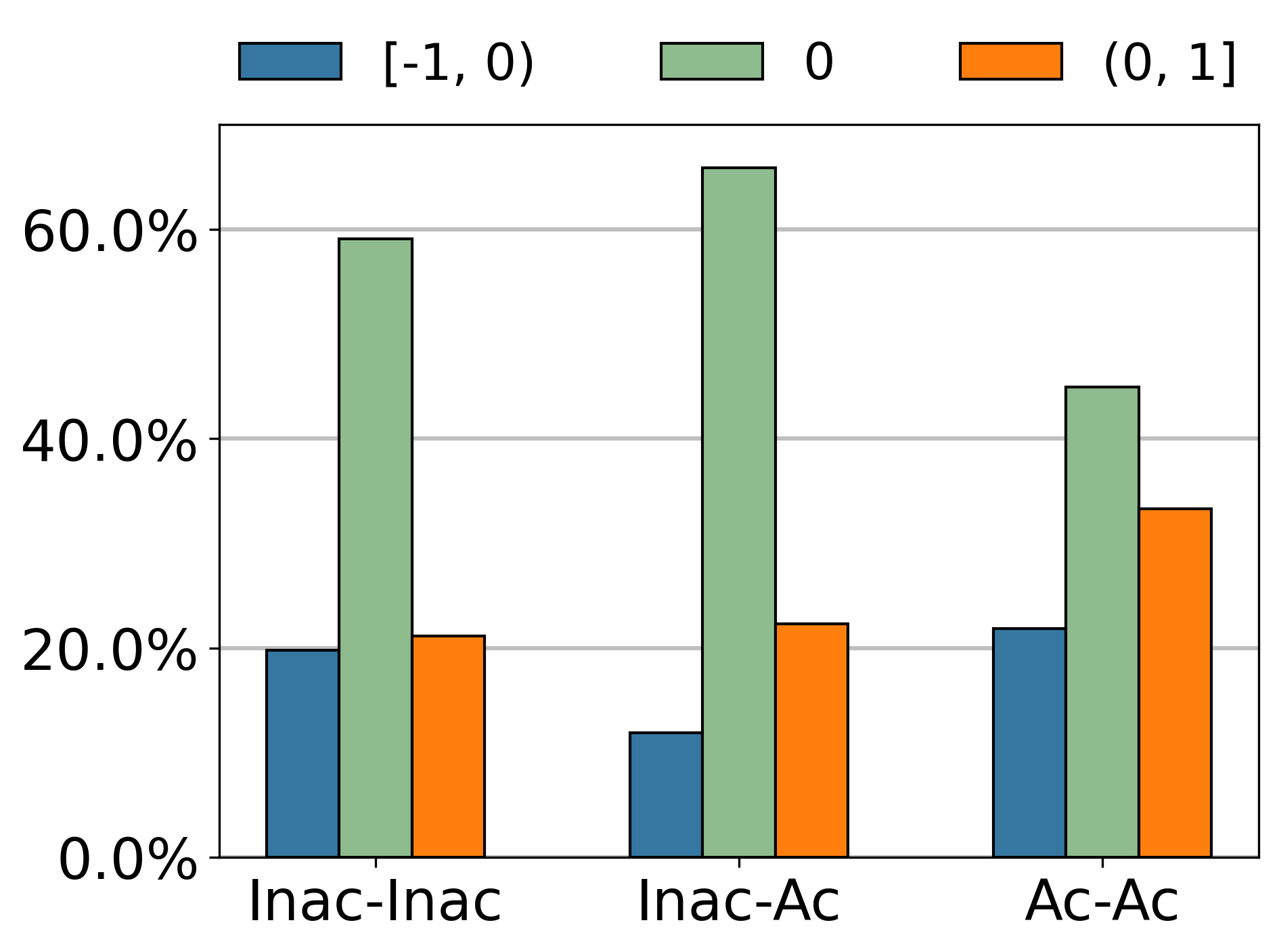}
    \end{minipage}
}
\caption{The industrial observations on social relations based on Taobao.}
\label{obe}
\end{figure*}
\setcounter{footnote}{0}
\section{Industrial Observations on Social Relation}
\label{observation}
In this section, we aim to investigate the impact of social relations on recommendation, especially for inactive users. We conduct our analysis based on real-world data from the Taobao e-commerce platform, part of Alibaba Group. Our observation includes over 100 million customers sampled uniformly from Taobao users. In this dataset, nearly 30\% of users are labeled as inactive, with less than 10 monthly online behavior records. We also include items purchased by these selected customers over two months (August \& September 2022)~\footnote{In this section, we focus on dynamic data from Taobao. Public datasets are not fine-grained enough (lack of timestamps and `inactive' definition). While the user population covered in this section is sufficiently large-scale, ensuring our observed phenomena are universally representative.}.


\textbf{Observation 1: Raw social relations generally have inferior quality.}
We analyze the quality of social relations based on different cases. These cases involve interactions between inactive users and active users ('Inac-Ac'), interactions between two inactive users ('Inac-Inac'), and interactions between two active users ('Ac-Ac'). Additionally, we randomly pair users and consider it as a separate case ('Rand'), with the number of random pairs equal to the total number of actual relations. To evaluate the quality of each relation, we measure the rate of user pairs that have a non-zero Jaccard index for the items they have purchased. A non-zero Jaccard index indicates that the two users have at least one common interest within the observation window.~\footnote{This evaluation is focused on recommendation accuracy, as a higher Jaccard index signifies a stronger correlation between users.} The results, shown in Fig.~\ref{obe}(a) as blue bars, indicate that for all three interaction types ('Inac-Inac', 'Inac-Ac', and 'Ac-Ac'), the rates of non-zero Jaccard index are no greater than 50\%. These rates only slightly exceed that of the 'Rand' case. Therefore, only a partial connection between users is capable of reflecting each other's purchase preferences, while the assistance from the remaining connections is limited. This limited assistance is considered as an indication of inferior quality.


\textbf{Observation 2: Social connections for inactive users is insufficient.} We examine the degree distributions of active users and inactive users in the social graph, as depicted in Fig.~\ref{obe} (b). The horizontal axis represents different intervals based on the social degrees of users. For instance, the interval (20, 50] includes users with more than 20 but no more than 50 social neighbors. The vertical axis represents the proportion of users falling into each interval.  we observe that in the intervals (0, 20], (20, 50], and (50, Max), the proportions of active users are higher than those of inactive users. However, there is a notable disparity when it comes to isolated inactive users, which account for 31.9\% of the total. In contrast, active users comprise only 6.2\% of this category. This suggests that, on average, inactive users have fewer social connections and possess less social information compared to active users.

\textbf{Observation 3: New-founded relations with active users bring more positive effect.} Our next observation focuses on the quality of newly established connections. Specifically, we examine the impact of these connections on user correlations. To do this, we calculate the difference in Jaccard indices between September and August for each pair of users who became connected during September. A higher positive difference indicates that the new connection has enhanced the correlations between these two users. We present these differences in Fig.~\ref{obe} (c). To our surprise, the analysis reveals that connecting with active users ('Inac-Ac' and 'Ac-Ac') tends to yield more positive gains. On the other hand, the correlation changes between connected inactive users ('Inac-Inac') appear to be more random. This observation suggests that establishing connections with active users holds greater value for recommendation purposes.

Based on the above observations, we realize that the quality of social relations is probably inferior. On this basis, inactive users only own insufficient neighbors, which aggravates the challenges to recommend for them. To address this issue, it is crucial to consider importing high-quality information from active users and leveraging it to benefit inactive users. By doing so, we can enhance the overall recommendation process and improve the situation for inactive users.
\setcounter{footnote}{0}
\section{Preliminary}
Let $U=\{u_1, u_2, \dots, u_m\}$ denotes the user set ($|U|=m$), and $I=\{i_1, i_2, \dots, i_n\}$ denotes the item set ($|I|=n$). Users and items are assigned with original features, denoted as $\bm{X}_U=[\bm{x}_{u_1}, \bm{x}_{u_2}, \dots, \bm{x}_{u_m}]\in\mathbb{R}^{m\times D_u}$ and $\bm{X}_I=[\bm{x}_{i_1}, \bm{x}_{i_2}, \dots, \bm{x}_{i_n}]\in\mathbb{R}^{n\times D_i}$. $I(u)$ represents the set of items purchased by user $u$, and $U(i)$ means the set of users interacting with item $i$. 
As for social recommendation, we utilize $N_u$ to denote user $u$'s neighbors in raw social graph.

\begin{definition}
    \textbf{Inactive/Active user.} In a fixed transaction period, user $u$ is defined as inactive user $u_{-}$ if $|I(u)|<\epsilon$, where $\epsilon$ is a pre-defined threshold according to business consensus. And the rest users are defined as active user, denoted as $u_{+}$. In this case, the user set $U=\{u_{+}, u_{-}\}$.
\end{definition}

\begin{definition}
    \textbf{Enhanced social recommendation via refining social graph.} The probability of item $i\in I$ is purchased by an user $u\in\{u_{+}, u_{-}\}$ is:
\begin{equation}
    \widehat{r}_{u_i} = \mathcal{F}(i\in I| u, I(u), N_{u}'),
\end{equation}
where the designed recommender $\mathcal{F}$ predicts the possibility as $\widehat{r}_{u_i}\in(0, 1)$. $N_{u}'$ is the refined social neighbor set by solving the problems about raw relations uncovered in industrial observations. In this paper, we primarily enhance the social recommendation for these inactive users $u_{-}$.
\end{definition}

\section{The Proposed Model}
\begin{figure*}[t]
  \centering
  \includegraphics[scale=0.25]{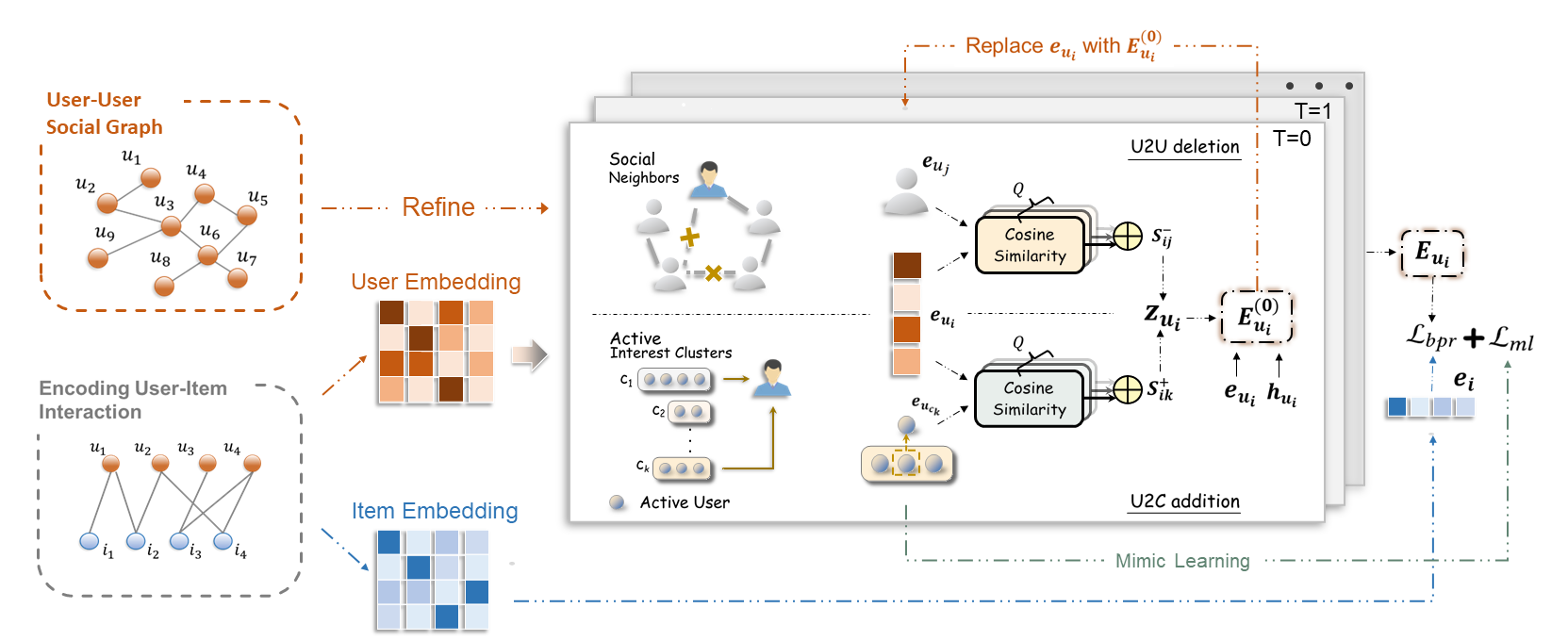}
  \caption{The overview of our proposed LSIR, including 3 components: 1)Encoding User-Item Interactions, 2)Graph Structure Learning on Social Graph(U2U deletion \& U2C addition) and 3)Mimic Learning.}
  \label{model}
\end{figure*}

In this section, we elaborate the proposed model LSIR, refining the social structure to improve the recommendation for inactive users. The overall architecture is shown in Fig.~\ref{model}, including encoding user-item interactions, refining user-user social graph and mimic learning. 
\subsection{Encoding User-Item Interactions}
To initially depict the characters of users and items, we focus on user-item interactions, and involve the collaborative filtering signals into their embeddings. Specifically, we need to firstly project features of users and items into a common latent vector space, since their features are composed in different ways. For an user $u$, we use a MLP with one hidden layer to transform his feature $\bm{x_u}$ into the common space:
\begin{equation}
\label{ori_u}
    \bm{h}_u = \bm{W}_U^{(1)}\sigma\left(\bm{W}_U^{(0)}\bm{x}_u+\bm{b}_U^{(0)}\right)+\bm{b}_U^{(1)},
\end{equation}
where $\bm{h}_u\in\mathbb{R}^{d\times 1}$ is the projected embedding of user $u$, $\sigma$ is PReLU non-linear function, and $\{\bm{W}_U^{(1)}, \bm{W}_U^{(0)}, \bm{b}_U^{(1)}, \bm{b}_U^{(0)}\}$ are learnable parameters, respectively. Meanwhile, the feature $\bm{x}_i$ of item $i$ can be mapped in a similar way:
\begin{equation}
    \bm{h}_i = \bm{W}_I^{(1)}\sigma\left(\bm{W}_I^{(0)}\bm{x}_i+\bm{b}_I^{(0)}\right)+\bm{b}_I^{(1)}.
\end{equation}
Next, we aim to enrich the user and item embeddings by aggregating interacted items and users, respectively. Considering potential degradation from further transformation and non-linear activation during aggregation~\cite{lightgcn}, we merely adopt the simple weighted sum aggregator:
\begin{equation}
\label{lgcn_1}
\bm{e}_u^{(k)} = \sum\limits_{i\in{I(u)}}\frac{\bm{e}_i^{(k-1)}}{\sqrt{|I(u)|}\sqrt{|U(i)|}}, \quad \bm{e}_i^{(k)} = \sum\limits_{u\in{U(i)}}\frac{\bm{e}_u^{(k-1)}}{\sqrt{|U(i)|}\sqrt{|I(u)|}},
\end{equation}
where $\bm{e}_u^{(k)}$ and $\bm{e}_i^{(k)}$ represent embeddings of users and items in the $k$th layer, $\bm{e}_u^{(0)}=\bm{h}_u$ and $\bm{e}_i^{(0)}=\bm{h}_i$. $|\cdot|$ is the length of set. After $K$ layers propagation, we further combine middle-layer results into the final embeddings of user $u$ and item $i$ under user-item interactions, denoted as $\bm{e}_u$ and $\bm{e}_i$:
\begin{equation}
\label{lgcn_2}
\bm{e}_u = \frac{1}{K}\sum\limits_{k=0}^K\bm{e}_u^{(k)},\quad \bm{e}_i = \frac{1}{K}\sum\limits_{k=0}^K\bm{e}_i^{(k)}.
\end{equation}

\subsection{Graph Structure Learning on Social Graph}
After encoding user-item interactions, the embedding $\bm{e_u}$ of inactive user $u_{-}$ still can not fully reflect his implicit interests, due to the limited consumption history. Thus, we need to supplement additional social information to boost the recommendation on $u_{-}$. Nevertheless, the raw social graph does not meet the requirements, pointed out in introduction. Hence, we refine it via two phases: U2U deletion and U2C addition. Next, we elaborate these two phases.

\paragraph{U2U deletion} Observation 1 tells us that only few raw social relations are valid. These effective edges should be carefully discriminated and retained, while other noisy connections should be filtered out. Therefore, for user $i$ and his one neighbor user $j$, we calculate the similarity between their embeddings to measure if the edge is effective. Without loss of generality, we adopt weighted cosine similarity~\cite{idgl} as their edge retention probability $S_{ij-}$:
\begin{equation}
\label{cos_1}
S_{ij-}=\cos(\bm{W}_n\cdot\bm{e}_{u_i}, \bm{W}_n\cdot\bm{e}_{u_j}),
\end{equation}
where $\bm{W}_n$ is learnable as a projection head to increase the flexibility of measurement. Furthermore, considering that there are multifaceted similarities between two users, we extend Eq.~\eqref{cos_1} to a multi-head version:
\begin{equation}
\label{cos_2}
s_{ij-}^{(q)}=\cos(\bm{W}_n^{(q)}\cdot\bm{e}_{u_i}, \bm{W}_n^{(q)}\cdot\bm{e}_{u_j}), \quad
S_{ij-}=\frac{1}{Q}\sum\limits_{q=1}^Qs_{ij-}^{(q)},
\end{equation}
where we average results from $Q$ aspects to get the comprehensive similarity. This multi-head extension can also stabilize the learning process~\cite{DBLP:conf/nips/VaswaniSPUJGKP17,gat}. With Eq.~\eqref{cos_2}, we can get retention probabilities between $u_i$ and all his neighbors. Next, we should consider how to adaptively select effective neighbors based on these probabilities for $u_i$. Please recall that inactive users absorb limited information from items (by definition) as well as social neighbors (Observation 2). Thus, the fewer items $u_i$ interacts with, the more social information we should maintain for him to avoid uninformative initialization. Based on this principle,  we devise the following tanh formula to calculate the deletion ratio for user $u_i$:
\begin{equation}
\label{p_dele}
p_{del}(|I(u_i)|)=\frac{\exp(|I(u_i)|/{r_1})-\exp(-|I(u_i)|/{r_1})}{\exp(|I(u_i)|/{r_1})+\exp(-|I(u_i)|/{r_1})},
\end{equation}
where $r_1$ is a hyper-parameter to control the smoothness. If $|I(u_i)|\rightarrow 0$, then $p_{del}(|I(u_i)|)\rightarrow 0$. We sort previous retention probabilities between $u_i$ and his neighbors in the descending order, and only retain top $1-p_{del}(|I(u_i)|)$ neighbors as the refined neighbor users set $N_{u_i}^U$.

\paragraph{U2C addition} \label{U2C Addition}
Enlightened by Observations 3, we realize that compared with randomly adding edges, linking inactive users to active ones is more efficient and reliable. However, active users still make up the vast majority (refer to Table~\ref{exp_stat}). If we directly connect inactive user to every active user, the refined social graph will be very dense. To tackle this problem, we creatively devise a \textit{U2C (user to cluster) addition} mechanism. Instead of employing all active users, we firstly mine interest clusters in active users (or active interest clusters), and then learn how to establish connections towards these clusters to reduce space complexity.


\textbf{Mining active interest clusters.} The core insight is to bundle active users into several clusters, and select representative active user in each cluster as anchor connecting to inactive users. The whole process is sketched as follows:
\begin{itemize}
\item Cluster items via their raw features, and obtain the clusters set $\mathcal{C}=\{c_1, c_2, \dots, c_l\}$, where each $c_i=\{i_1^{c_i}, i_2^{c_i}, \dots, i_{|c_i|}^{c_i}\}$ represents one specific interest.
\item For active user $u_{+}$, we calculate Jaccard index between $I(u_{+})$ and $c_i$, denoted as $\mathcal{J}(I(u_{+}), c_i)$, and find the best-match cluster $\mathcal{C}(u_{+})$:
\begin{equation}
\label{cu_{+}}
\mathcal{C}(u_{+})=\arg\max\limits_{c_i}\mathcal{J}(I(u_{+}), c_i),
\end{equation}
\item For cluster $c_i$, we construct $\mathbb{U}(c_i)=\{u_{+}|\mathcal{C}(u_{+})=c_i\}$ via last step. In $\mathbb{U}(c_i)$, we choose the user $u_{c_i}$ who owns the maximal Jaccard index with cluster $c_i$.
\item Group these chosen users $U_C=\{u_{c_1}, u_{c_2}, \dots, u_{c_l}\}$ to represent each corresponding item cluster.
\end{itemize}
Please note the above mining process only involves a hyper-parameter $l$, the cluster number. Next, we again resort to multi-head cosine similarity to define the probability $S_{i{k}+}$ of connection between user $u_i$ and one anchor user $u_{c_k}\in U_C$:
\begin{equation}
\label{cos_3}
s_{i{k}+}^{(q)}=\cos(\bm{W}_n^{(q)}\cdot\bm{e}_{u_i}, \bm{W}_c^{(q)}\cdot\bm{e}_{u_{c_k}}), \quad
S_{ik+}=\frac{1}{Q}\sum\limits_{q=1}^Qs_{i{k}+}^{(q)},
\end{equation}
where $\bm{W}_n^{(q)}$ is shared with that in Eq.~\eqref{cos_2}, but a new learnable matrix $\bm{W}_c^{(q)}$ is utilized to project anchor users specifically. Then, we still need to adaptively select the addition ratio for each user. Due to the inadequate social neighbors for inactive users (Observation 2), more information from high-quality active users should be involved into them as a supplementary. Thus, the addition ratio for user $u_i$ is:
\begin{equation}
\label{p_add}
p_{add}(|I(u_i)|)=\frac{1}{1+\exp (|I(u_i)|/r_2)},
\end{equation}
where $r_2$ is also a hyper-parameter similar to $r_1$. The less $|I(u_i)|$ is, the larger $p_{add}(|I(u_i)|)$ is. For $u_i$, we select top $p_{add}(|I(u_i)|)$ anchor users with maximal $S_{i{k}+}$ as set $N_{u_i}^C$.

Now, we have got the refined neighbor users set $N_{u_i}^U$ and candidate anchor users set $N_{u_i}^C$ for $u_i$, then we fuse them into $\bm{z_{u_i}}$ under refined user-user social graph: 
\begin{equation}
\label{uu_u}
\bm{z_{u_i}}=\alpha\left(\sum\limits_{j\in N_{u_i}^U} S_{ij-}\cdot\bm{e}_{u_j}+\sum\limits_{k\in N_{u_i}^C}S_{i{k}+}\cdot\bm{e}_{u_{c_k}}\right)+(1-\alpha)\bm{e}_{u_i},
\end{equation}
where $\alpha$ is a balance parameter. Combining Eq.~\eqref{ori_u},~\eqref{lgcn_2} and~\eqref{uu_u}, we utilize skip-connection to fuse these three user embeddings to get $\bm{E}_u^{(0)}$ at the first iteration:
\begin{equation}
\label{final_u}
\bm{E}_u^{(0)} = \bm{W}_f\cdot [\bm{h}_u||\bm{e}_u||\bm{z}_u]+\bm{b}_f,
\end{equation}
where $\bm{W}_f$ and $\bm{b}_f$ are learnable parameters. Then, we replace $\bm{e}_u$ with $\bm{E}_u^{(0)}$, and redeploy U2U deletion and U2C addition to seek for a better social structure. After such $T$ iterations, we get user final embedding $\bm{E}_u$.

Please note that although we mainly describe U2U deletion and U2C addition from the perspective of inactive users, this refinement process is also employed for active ones. Considering their large degree in social neighbors, we remove a large amount of useless social neighbors and very prudently involve new information for them according to Eq.~\eqref{p_dele} and~\eqref{p_add}. Therefore, the performance on active users can be guaranteed with improving inactive users.

To optimize the LSIR, BPR loss~\cite{bpr} is adopted as following:
\begin{equation}
\label{bpr_loss}
\mathcal{L}_{bpr}=\sum\limits_{(u, i, i')\in \mathcal{O}} -\log\sigma(\widehat{r}_{ui}-\widehat{r}_{ui'})+\lambda||\Theta||^2,
\end{equation}
where $\Theta$ denotes the parameters of LSIR, $\widehat{r}_{ui}=\bm{E}_u^\top\bm{e}_i$ is the predicted score of $u$ on $i$, and $\sigma$ is sigmoid. Each triple instance in $\mathcal{O}$ includes target $u$, a ground-truth item $i$, and a negative item $i'$ randomly chosen from unobserved items. 

\subsection{Mimic Learning}
\label{Mimic Learning}
Although we have refined the raw user-user social graph, we notice that in U2C addition part, it is still difficult to match inactive users with suitable clusters. This is because in Eq.~\eqref{cos_3}, only $\bm{e}_u$ (got from Eq.~\eqref{lgcn_2}) is utilized to measure the similarity, which only absorbs limited items for inactive users that cannot support them to connect towards true clusters. To solve this problem, we manipulate LSIR to know how to map users to corresponding clusters via the proposed mimic learning mechanism. Our intention is to generate pseudo inactive user embeddings $\bm{\hat{e}}_{u_{+}}$ for active user $u_{+}$. In this case, LSIR is expected to still match the changed $\bm{\hat{e}}_{u_{+}}$ to the real best-match $\mathcal{C}(u_{+})$, given in Eq.~\eqref{cu_{+}}.

MixUp~\cite{mixup} is proposed to efficiently improve results in supervised learning by adding arbitrary two samples to create a new one. MixGCF~\cite{mixgcf} introduces this strategy into graph-based recommendation, who involves positive samples into negative ones to generate harder negatives. Inspired by them, we inject inactive information into active embeddings, we call this method \textbf{Inactive Mixture}. Firstly, for each active user $u_{+}$, we randomly select $\eta$ inactive users in one batch, and combine them with $u_{+}$ with weight $\beta\in(0,1)$:
\begin{equation}
\label{im}
\bm{\hat{e}}_{u_{+}}=\beta\bm{e}_{u_{+}}+(1-\beta)\bm{e}_{u_{-}},
\end{equation}
where $\bm{e}_{u_{-}}$ is embedding of one selected inactive user.

With the generated pseudo inactive user embeddings $\bm{\hat{e}}_{u_{+}}$, we manipulate encoder to make $\bm{\hat{e}}_{u_{+}}$ close to the anchor user who represents cluster $\mathcal{C}(u_{+})$. To achieve this target, we adopt contrastive InfoNCE loss~\cite{cpc}:
\begin{equation}
\label{mimic_loss}
\mathcal{L}_{ml}=-\sum\limits_{u_{+}}\log\frac{\exp(\cos(\bm{\hat{e}}_{u_{+}}, \bm{e}_{u_{c_j}})/\tau)}{\sum_{k\neq j}\exp(\cos(\bm{\hat{e}}_{u_{+}}, \bm{e}_{u_{c_k}})/\tau)},
\end{equation}
where we assume $\mathcal{C}(u_{+})=c_j$, and $\tau$ is a temperature parameter. With the optimization of Eq.~\eqref{mimic_loss}, the encoder is trained to match pseudo inactive $\bm{\hat{e}}_{u_{+}}$ to the real $\mathcal{C}(u_{+})$ and keep away from other clusters, even though $u_{+}$ has attempted to mimic other inactive users.


Combined with Eq.~\eqref{bpr_loss}, the overall loss is:
\begin{equation}
\label{omega_loss}
\mathcal{L}_{\Omega}=\mathcal{L}_{bpr}+\xi\cdot\mathcal{L}_{ml},
\end{equation}
where $\xi$ is a combination coefficient. We can optimize the proposed LSIR via back propagation with stochastic gradient descent. 

\subsection{Complexity Analysis}
The complexity of each iteration inside LSIR mainly involves three parts: encoding user-item interactions, refining user-user social graph and mimic learning. For the first part, the propagation between users and items consumes $\mathcal{O}(|\varepsilon|dK)$, where $\varepsilon$ is the set of all user-item interactions, and $d$ is the dimension of embedding. As for the refining process, we measure $Q$-head cosine similarity between target user and his neighbors or clusters. So, the complexity of this part is $\mathcal{O}(Q(|\zeta|+ml)d)$, where $\zeta$ is the set of all social relations and $m$ is the number of users, and $l$ is the number of active interest clusters. The cost of mimic learning is $\mathcal{O}(|B_+|\eta ld)$, where $B_+$ is the set of active users in one batch, and $\eta$ is the number of selected inactive users. In summary, the complexity of LSIR is around $\mathcal{O}(|\varepsilon|dK+Q(|\zeta|+ml)d+|B_+|\eta ld)$, which is linear to $|\varepsilon|$ and $|\zeta|$. On one hand, the complexities of previous GNNs-based social recommendation models~\cite{mhcn,diffnet} are also linear to $|\varepsilon|$ and $|\zeta|$, which indicates that we do not drastically increase the model complexity. On the other hand, the complexities of traditional GSL based methods~\cite{lds,prognn} are $\mathcal{O}(|\varepsilon|dK+m^2d)$, which is much higher than that of our proposed LSIR, considering $|\zeta|\ll m^2$.
\section{Experiments}

In this section, we conduct experiments to evaluate the performance of LSIR. Specifically, we aim to answer the following research questions:

\begin{itemize}

\item \textbf{RQ1}: How does LSIR perform compared to state-of-the-art baselines?
\item \textbf{RQ2}: How does graph structure learning contribute to model performance?
\item \textbf{RQ3}: How do key hyper-parameters affect the model?
\end{itemize}
\subsection{Experimental Setup}


\begin{table*}[t]
  \centering
  \caption{The statistics of four datasets.}
  \resizebox{0.8\textwidth}{!}{
  \begin{tabular}{c|c|c|c|c|c|c}
    \bottomrule
    Dataset & \#User & \#Inactive User & \#User-User & \#Item & \#User-Item & Inactive Cutoff\\
    \hline
    Flickr & 8,358 &2,386&187,273&82,120&327,815& $\leq 2$ \\
    Yelp & 17,237 &5,445&143,765&38,342&204,448& $\leq 3$ \\
    AliData $\uppercase\expandafter{\romannumeral1}$&40,428&12,071&323,608&115,502&1,016,122& $\leq 7$ \\
    AliData $\uppercase\expandafter{\romannumeral2}$&103,440&23,335&984,555&2,417,818&12,095,268& $\leq 20$ \\
    \bottomrule
    \end{tabular}}
  \label{exp_stat}
\end{table*}

\textbf{Datasets.} We employ four datasets, including two publicly available datasets~\cite{diffnet++} (i.e. Flickr, Yelp) and two real industrial datasets (i.e. AliData $\uppercase\expandafter{\romannumeral1}$, AliData $\uppercase\expandafter{\romannumeral2}$). AliData $\uppercase\expandafter{\romannumeral1}$ represents a smaller-scale market with specific commodity categories and customer groups, where users have more sparse and concentrated behaviors. Contrarily, AliData $\uppercase\expandafter{\romannumeral2}$ covers a broader range of commodities and serves diverse customers with different shopping needs, where interactions are denser. To provide transparency regarding our selection of inactive users, we present the cutoff criteria for inactive user behavior of each dataset in Table~\ref{exp_stat}.

\noindent\textbf{Baselines.} We compare LSIR with three categories of baselines, including general recommendation methods \{BPR~\cite{bpr}, LightGCN~\cite{lightgcn}, SimGCL~\cite{simgcl}\}, social recommendation methods \{SBPR~\cite{sbpr}, DiffNet++~\cite{diffnet++}, $S^2$-MHCN~\cite{mhcn}, SEPT~\cite{sept}\} and cold-start recommendation methods \{MeLU~\cite{melu}, TaNP~\cite{tanp}\}. 

\noindent\textbf{Metrics.} We focus on performing Top-K recommendation, where $K\in\{10, 20\}$. we choose two ranking based metrics \textit{Normalized Discounted Cumulative Gain} (NDCG)~\cite{DBLP:journals/tois/DeshpandeK04} and \textit{Hit Ratio} (HR)~\cite{DBLP:journals/tois/DeshpandeK04}, and one relevancy-based metric \textit{Precision} (PR)~\cite{mhcn}. For each user, we randomly sample 1000 unrelated items as negative samples, and combine them with the ground-truth items in the ranking process to select top-k items for the final evaluation, which follows DiffNet++~\cite{diffnet++}.

\noindent\textbf{Implementation Details.} For a fair comparison, we set the dimension of embeddings as 64, and follow the settings in the original papers with carefully tuning. For LSIR, we use Glorot initialization~\cite{glorot2010understanding} and Adam~\cite{adam} optimizer. We tune the learning rate of $\Theta$ and $\Omega$ from 1e-4 to 1e-2. For temperature coefficients $r_1$ and $r_2$, we test ranging from \{2, 10, 50, 100\}. We search the number of attention heads from \{1, 2, 3\}. Finally, we tune $\xi$ in \{0.1, 0.01\}. 
\begin{table*}[t]
  \caption{Quantitative results on Top-10 and Top-20 recommendation for inactive users. (bold: best; underline: runner-up; `-': Out-Of-Memory on 32GB GPU; `Improv.' denotes the relative improvements over the best baseline)}
  \centering
  \resizebox{\textwidth}{!}{
  \begin{tabular}{c|c|c|ccc|cccc|cc|c|c}
    \bottomrule
    Datasets & Metric &Top-K & BPR & LightGCN & SimGCL & SBPR & DiffNet++ & $S^2$-MHCN & SEPT & MeLU & TaNP&\textbf{LSIR}&\textbf{Improv.}\\
    \bottomrule
    \multirow{6}{*}{Flickr}&
    \multirow{2}{*}{NDCG}
    &10&0.0503	&0.0797	&0.0648	&0.0659	&\underline{0.1251}	&0.0990	&0.0455	&0.0660	&0.0781	&\textbf{0.2872}	&\textbf{129.58\%}\\
    &&20&0.0630	&0.1042	&0.0801	&0.0808	&\underline{0.1470}	&0.1137	&0.0561	&0.0994	&0.1068	&\textbf{0.3253}	&\textbf{121.29\%}\\
    \cline{2-14}
    &\multirow{2}{*}{HR}
    &10&0.0731	&0.1350	&0.0951	&0.0997	&\underline{0.1752}	&0.1413	&0.0675	&0.1154	&0.1391	&\textbf{0.3927}	&\textbf{124.14\%}\\
    &&20&0.1112	&0.2094	&0.1444	&0.1465	&\underline{0.2462}	&0.2000	&0.1028	&0.2045	&0.2176	&\textbf{0.5164}	&\textbf{109.75\%}\\
    \cline{2-14}
    &\multirow{2}{*}{PR}
    &10&0.0146	&0.0270	&0.0190	&0.0199	&\underline{0.0350}	&0.0283	&0.0135	&0.0231	&0.0273	&\textbf{0.0785}	&\textbf{124.29\%}\\
    &&20&0.0111	&0.0209	&0.0144	&0.0147	&\underline{0.0246}	&0.0200	&0.0103	&0.0204	&0.0217	&\textbf{0.0516}	&\textbf{109.76\%}\\
    
    \hline
    \multirow{6}{*}{Yelp}&
    \multirow{2}{*}{NDCG}
     &10&0.1437	&0.1636	&0.1847	&0.1678	&\underline{0.2261}	&0.1845	&0.1480	&0.1485	&0.1562	&\textbf{0.2669}	&\textbf{18.05\%}\\
    &&20&0.1725	&0.1945	&0.2152	&0.2002	&\underline{0.2605}	&0.2151	&0.1834	&0.1969	&0.2126	&\textbf{0.3038}	&\textbf{16.62\%}\\
    \cline{2-14}
    &\multirow{2}{*}{HR}
    &10&0.2185	&0.2439	&0.2709	&0.2525	&\underline{0.3288}	&0.2742	&0.2274	&0.2492	&0.2642	&\textbf{0.3804}	&\textbf{15.69\%}\\
    &&20&0.3103	&0.3409	&0.3720	&0.3555	&\underline{0.4495}	&0.3754	&0.3357	&0.3971	&0.4085	&\textbf{0.4997}	&\textbf{11.17\%}\\
    \cline{2-14}
    &\multirow{2}{*}{PR}
    &10&0.0437	&0.0487	&0.0542	&0.0505	&\underline{0.0658}	&0.0548	&0.0455	&0.0498	&0.0531	&\textbf{0.0761}	&\textbf{15.65\%}\\
    &&20&0.0310	&0.0304	&0.0372	&0.0356	&\underline{0.0450}	&0.0375	&0.0336	&0.0397	&0.0408	&\textbf{0.0500}	&\textbf{11.11\%}\\

    \hline
    \multirow{6}{*}{AliData $\uppercase\expandafter{\romannumeral1}$}&
    \multirow{2}{*}{NDCG}
    &10&0.1634	&0.1629	&0.1815	&0.1644	&0.1793	&\underline{0.1817}	&0.1650	&	0.1648&	0.1696&\textbf{0.2018}	&\textbf{11.06\%}\\
    &&20&0.1817	&0.1766	&0.1931	&0.1812	&0.1990	&\underline{0.2019}	&0.1798	&	0.1784&	0.1810&\textbf{0.2214}	&\textbf{9.66\%}\\
    \cline{2-14}
    &\multirow{2}{*}{HR}
    &10&0.2201	&0.2129	&0.2224	&0.2169	&\underline{0.2515}	&0.2509	&0.2122	&	0.2219&	0.2223&\textbf{0.2748}	&\textbf{9.26\%}\\
    &&20&0.2859	&0.2622	&0.2638	&0.2791	&0.3223	&\underline{0.3227}	&0.2655	&	0.2660&	0.2724&\textbf{0.3530}	&\textbf{9.39\%}\\
    \cline{2-14}
    &\multirow{2}{*}{PR}
    &10&0.0330	&0.0301	&0.0333	&0.0325	&\underline{0.0377}	&0.0376	&0.0318	&	0.0317&	0.0324&\textbf{0.0412}	&\textbf{9.28\%}\\
    &&20&0.0214	&0.0187	&0.0198	&0.0209	&0.0241	&\underline{0.0242}	&0.0199&	0.0202&		0.0204&\textbf{0.0265}	&\textbf{9.50\%}\\
    
    \hline
    \multirow{6}{*}{AliData $\uppercase\expandafter{\romannumeral2}$}&
    \multirow{2}{*}{NDCG}
    &10&0.1126&\underline{0.1812}	&0.1757	&	0.1327&-	&	-&0.1796	&	0.1777&	0.1808& \textbf{0.2060}&	\textbf{13.69\%}\\
    &&20&0.1251	&	\underline{0.1903}&	0.1821&	0.1414&	-&-	&0.1897	&	0.1840&0.1866	&	\textbf{0.2208}&\textbf{16.03\%}\\
    \cline{2-14}
    &\multirow{2}{*}{HR}
    &10&	0.1191&0.1638	&0.1554	&0.1368	&-	&-	&\underline{0.1688}	&	0.1610&0.1612	&	\textbf{0.2092}&\textbf{23.93\%}\\
    &&20&0.1575	&0.1955	&0.1787	&0.1664	&-	&	-&\underline{0.2037}&	0.1812&	0.1855&	\textbf{0.2663}&	\textbf{30.73\%}\\
    \cline{2-14}
    &\multirow{2}{*}{PR}
    &10&0.0329	&0.0452	&0.0430	&0.0378	&-	&-	&\underline{0.0467}&	0.0435&	0.0444&	\textbf{0.0578}&\textbf{23.77\%}\\
    &&20&0.0218	&0.0270	&0.0247	&0.0230	&-	&	-&\underline{0.0282}&	0.0260&	0.0262&		\textbf{0.0368}&\textbf{30.50\%}\\
    
    \bottomrule
  \end{tabular}}
  \label{inac_results}
\end{table*}

\begin{table*}[t]
  \caption{The NDCG@10 performances on active and overall users got from social recommenders. (bold: best; underline: runner-up; `-': Out-Of-Memory on 32GB GPU)}
\label{asdads}
\centering
  \resizebox{0.75\textwidth}{!}{
  \begin{tabular}{c|cc|cc|cc|cc}
    \hline
    Datasets & \multicolumn{2}{c|}{Flickr} & \multicolumn{2}{c|}{Yelp} & \multicolumn{2}{c|}{AliData $\uppercase\expandafter{\romannumeral1}$} & \multicolumn{2}{c}{AliData $\uppercase\expandafter{\romannumeral2}$} \\
    \hline
     Methods & Active & Overall & Active & Overall & Active & Overall & Active & Overall\\
    \hline
    SBPR&0.0728&0.0709&0.1687&0.1685&0.2128&0.1984&0.2964&0.2595\\
    DiffNet++&\underline{0.1011}&\underline{0.1080}&\underline{0.2193}&\underline{0.2213}&0.2138&0.2035&-&-\\
    $S^2$-MHCN&0.0918&0.0939&0.1991&0.1950&0.2274&0.2138&-&-\\
    SEPT&0.0593&0.0554&0.1701&0.1638&\textbf{0.2362}&\underline{0.2150}&\textbf{0.3415}&\textbf{0.3050}\\
    \hline    \textbf{LSIR}&\textbf{0.1010}&\textbf{0.1542}&\textbf{0.2317}&\textbf{0.2418}&\underline{0.2281}&\textbf{0.2203}&\underline{0.3335}&\underline{0.3048}\\
    \hline
  \end{tabular}}
  
\end{table*}

\setcounter{footnote}{0}
\subsection{Overall Recommendation Performance(RQ1)}
\label{saeaseaseae}
In this subsection, we aim to validate the effectiveness of LSIR on recommendation. Specifically, we report the performances on Top-10/20 recommendation for inactive users in Table~\ref{inac_results}, and NDCG@10 results for active and overall users in Table~\ref{asdads}, where we focus on social recommenders. The conclusions are as follows:
\begin{itemize}
    \item \textbf{LSIR consistently outperforms other baselines for inactive users.} Especially, LSIR achieves a significant improvement of 129.58\% on NDCG of Flickr and 30.73\% on HR of AliData $\uppercase\expandafter{\romannumeral2}$. This demonstrates the effectiveness of LSIR for inactive users. Besides, our model can successfully run on large-scale AliData $\uppercase\expandafter{\romannumeral2}$, while DiffNet++ and $S^2$-MHCN fail, showing the scalability of LSIR.
    \item \textbf{The results on Flickr stand out.} The reason is related to the average social neighbors per user. According to Table~\ref{exp_stat}, the order follows Flickr (22.41) $\gg$ AliData $\uppercase\expandafter{\romannumeral2}$ (9.52) $>$ Yelp (8.34) $>$ AliData $\uppercase\expandafter{\romannumeral1}$ (8.00), which is in agreement with the order of performance improvements. Thus, higher number of neighbors leaves a larger room for LSIR to boost social relations.
    \item \textbf{Cold-start recommenders show weak competence in all the cases.} There are two possible causes. Firstly, these two cold-start models(MeLU and TaNP) are based on MAML rather than GNN, so superiority from message passing is neglected. Secondly, they do not absorb the useful knowledge from social relations, and only depend on user-item interactions.
    \item \textbf{LSIR is competitive constantly in both active and overall user cases.} As shown in Fig.~\ref{asdads}, LSIR consistently achieves the best or second-best performance across all users in both academic and industrial datasets. This indicates that the proposed refinement process will not sacrifice the performance on active and overall users. LSIR successfully strikes a balance between enhancing recommendation quality for inactive users while maintaining or even improving performance for active and overall users.
\end{itemize}

\subsection{Effects of Graph Structure Learning(RQ2)}

\subsubsection{Analysis of GSL variants}

\begin{wraptable}[12]{r}{0.5\textwidth}
  \centering
  \caption{Test on the variants of GSL. (`-': Out-Of-Memory on 32GB GPU)}
  \resizebox{0.45\textwidth}{!}{
  \begin{tabular}{c|cc|cc}
    \bottomrule
    Datasets & \multicolumn{2}{|c|}{Yelp} & \multicolumn{2}{|c}{AliData $\uppercase\expandafter{\romannumeral1}$}\\
    \bottomrule
    Metric & N@10 & N@20  & N@10 & N@20 \\
    \hline
    LSIR+UU & 0.2211 &0.2551 & 0.1984 &0.2186 \\
    LSIR+IDGL&0.2066&0.2404&0.1883&0.2062\\
    LSIR+GAUG&0.1974&0.2317&-&-\\
    \hline 
    LSIR w/o U2C&0.2590&0.2903&0.1991&0.2187\\
    LSIR w/o U2U&0.2622&0.2994&0.2009&0.2200\\   
    \hline  \textbf{LSIR}&\textbf{0.2669}&\textbf{0.3038}&\textbf{0.2018}&\textbf{0.2214}\\
    \bottomrule
\end{tabular}}
  \label{variants}
\end{wraptable}
To evaluate the effectiveness of designed GSL component in LSIR, we replace this part with two other GSL methods IDGL~\cite{idgl} and GAUG~\cite{gaug}, denoted as LSIR+IDGL and LSIR+GAUG. In addition, we have a baseline variant, LSIR+UU, where the original social graph remains unchanged. We compare these variants with our model, and report $\textit{NDCG@10}$ and $\textit{NDCG@20}$ on inactive users on Yelp and AliData $\uppercase\expandafter{\romannumeral1}$ in Table~\ref{variants}. From the table, we observe that LSIR+GAUG fails to run on AliData $\uppercase\expandafter{\romannumeral1}$ due to its high space complexity. LSIR+IDGL performs even worse than LSIR+UU because it neglects some important users in its random anchors. Additionally, we explore the effectiveness of U2C addition and U2U deletion individually. The results, as shown in the Table~\ref{variants}, indicate that the absence of the U2C addition component leads to a larger drop in performance compared to the absence of the U2U deletion component. This indicates that for inactive users, high-quality supplement is more valuable than deleting limited neighbors.

\subsubsection{Analysis of mimic learning}
\begin{wrapfigure}[11]{r}{0.6\textwidth}
\centering
    \subfigure[Flickr]
{
    \begin{minipage}[b]{.25\textwidth}
        \centering
        \includegraphics[width=\textwidth]{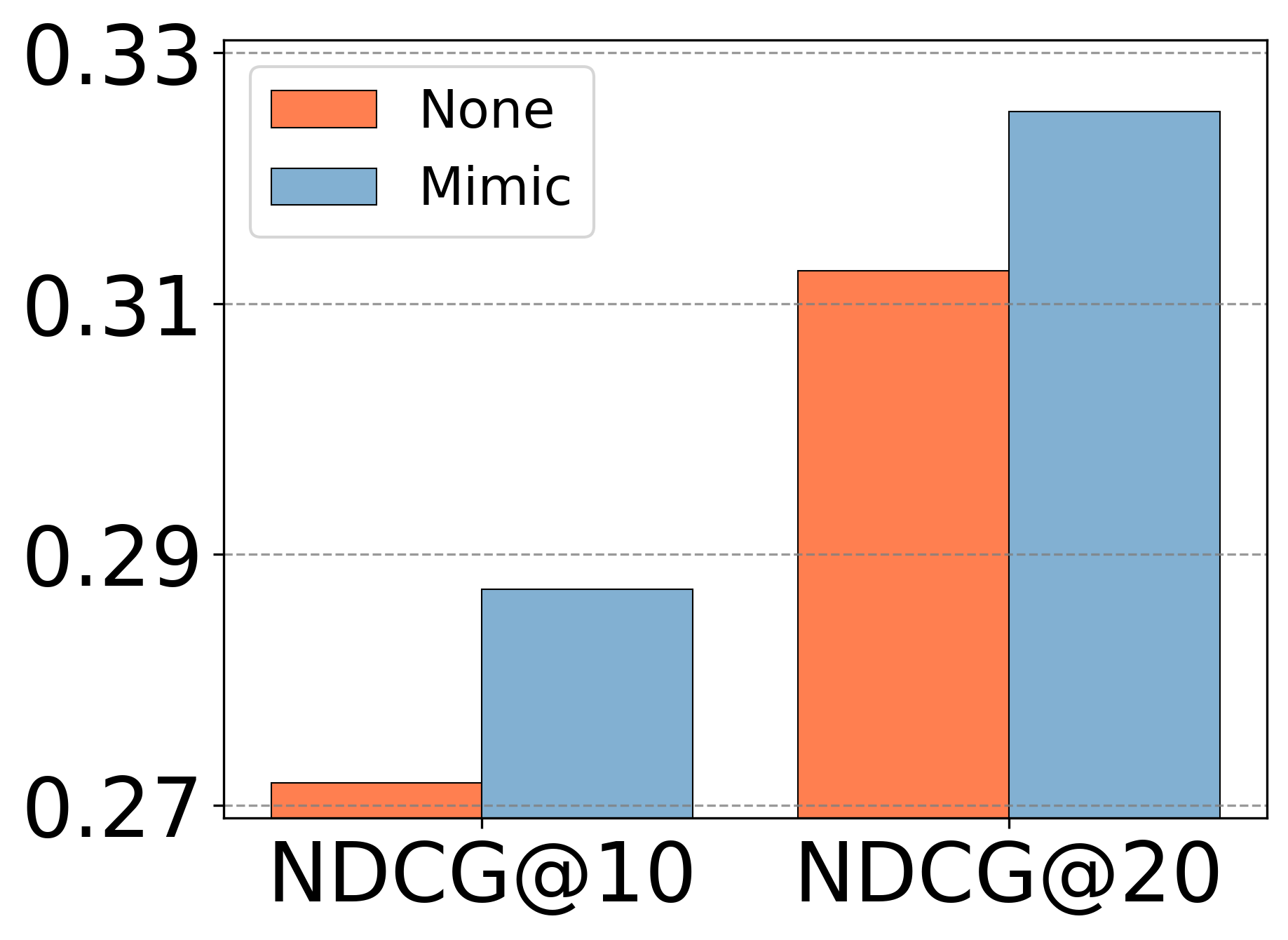}
    \end{minipage}
}
\subfigure[Yelp]
{
 	\begin{minipage}[b]{.25\textwidth}
        \centering
        \includegraphics[width=\textwidth]{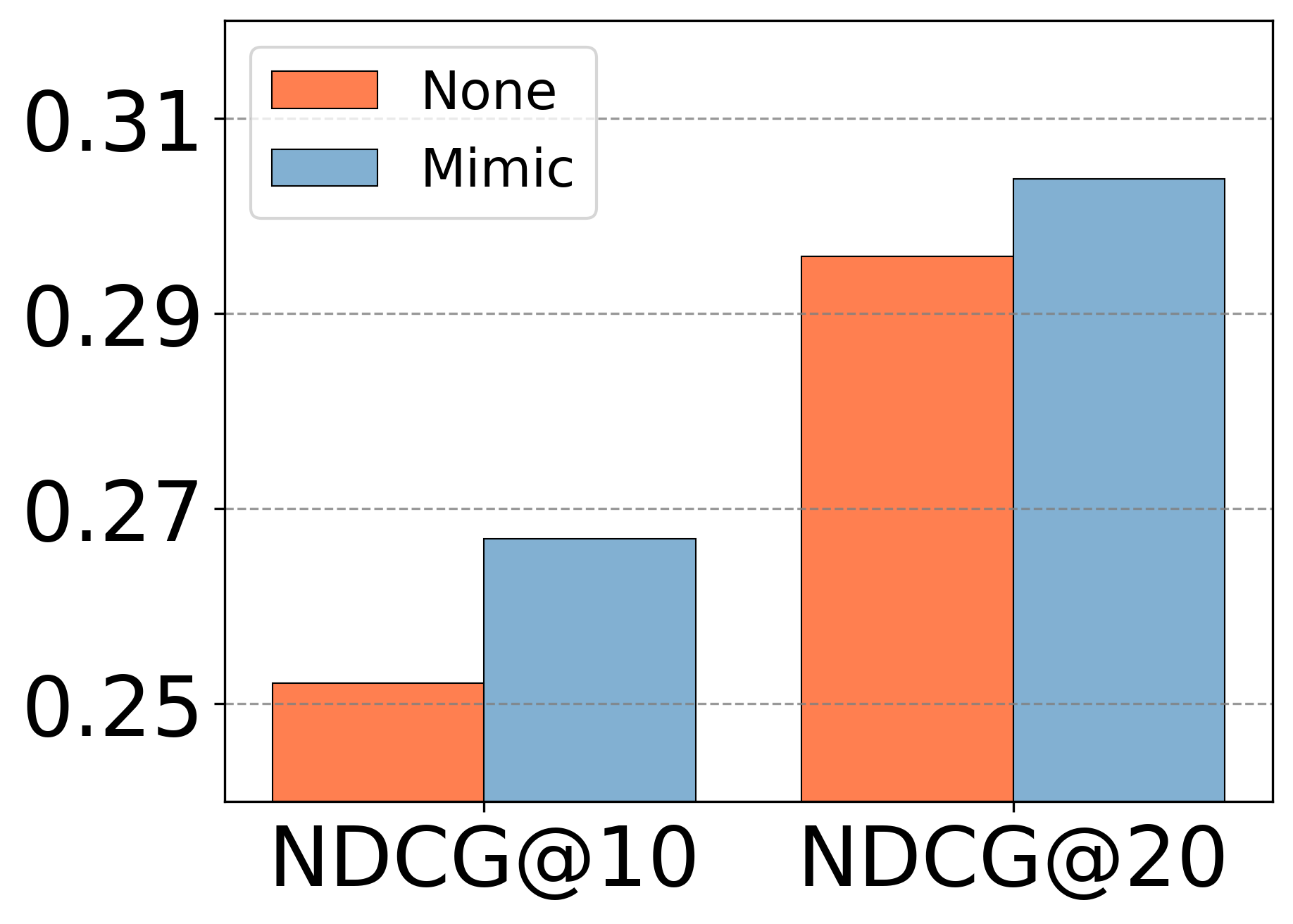}
    \end{minipage}
}
\caption{Test on the Mimic Learning.}
\label{mimimi}
\end{wrapfigure}
We test the effectiveness of mimic learning to enhance GSL. The results are given in Fig.~\ref{mimimi}, where Flickr and Yelp results are reported. In the figure, `None' means no mimic learning. As shown in figure, the mimic learning performs better than `None', with relative improvements around 6\% on NDCG@10, indicating the effectiveness of this component.

\subsection{Effects of Hyper-parameters(RQ3)}

\subsubsection{Analysis of ratio of inactive user} 

\begin{wrapfigure}[12]{r}{0.6\textwidth}
\centering
    \subfigure[Top-10]
{
    \begin{minipage}[b]{.25\textwidth}
        \centering
        \includegraphics[width=\textwidth]{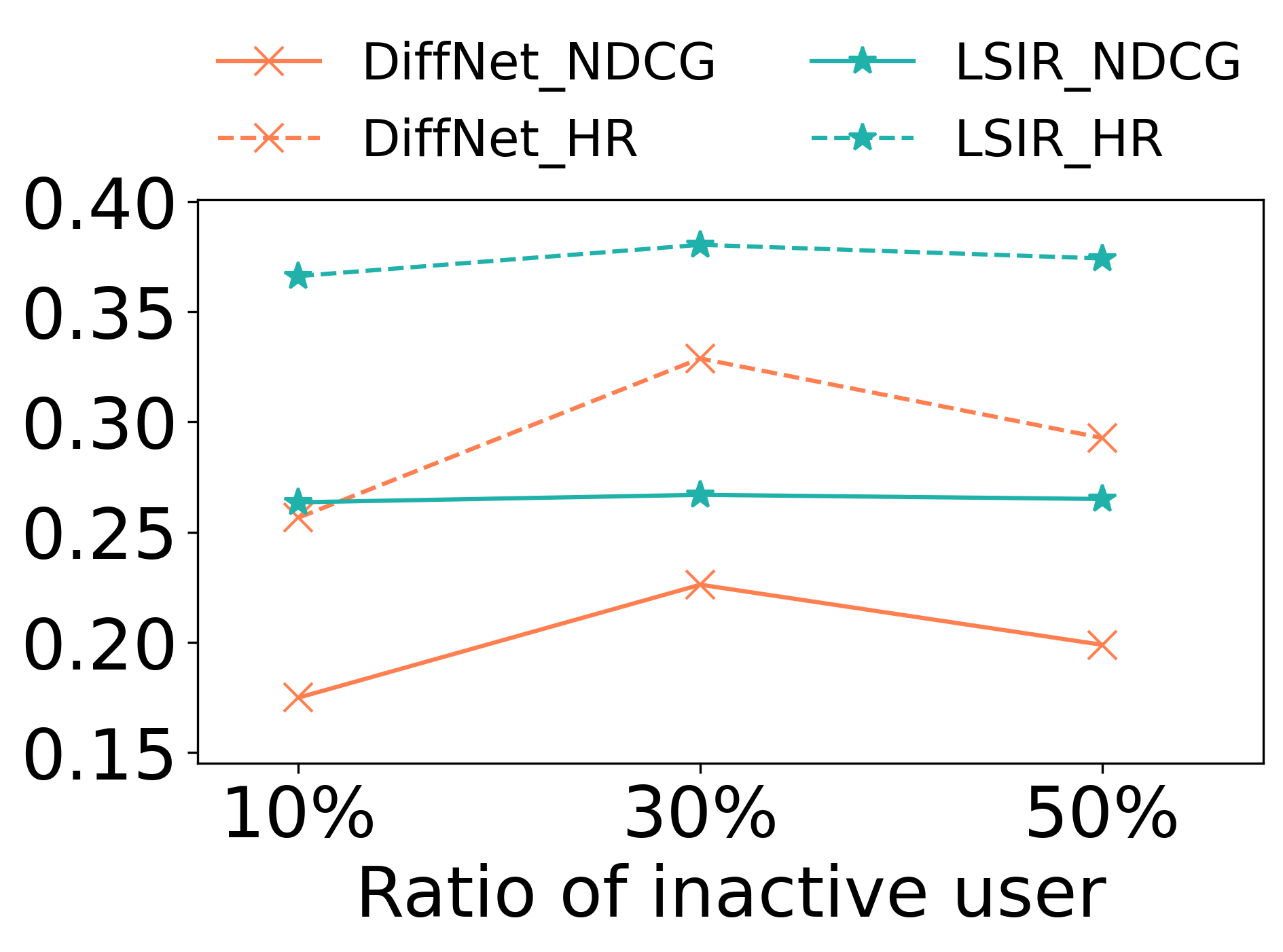}
    \end{minipage}
}
\subfigure[Top-20]
{
 	\begin{minipage}[b]{.25\textwidth}
        \centering
        \includegraphics[width=\textwidth]{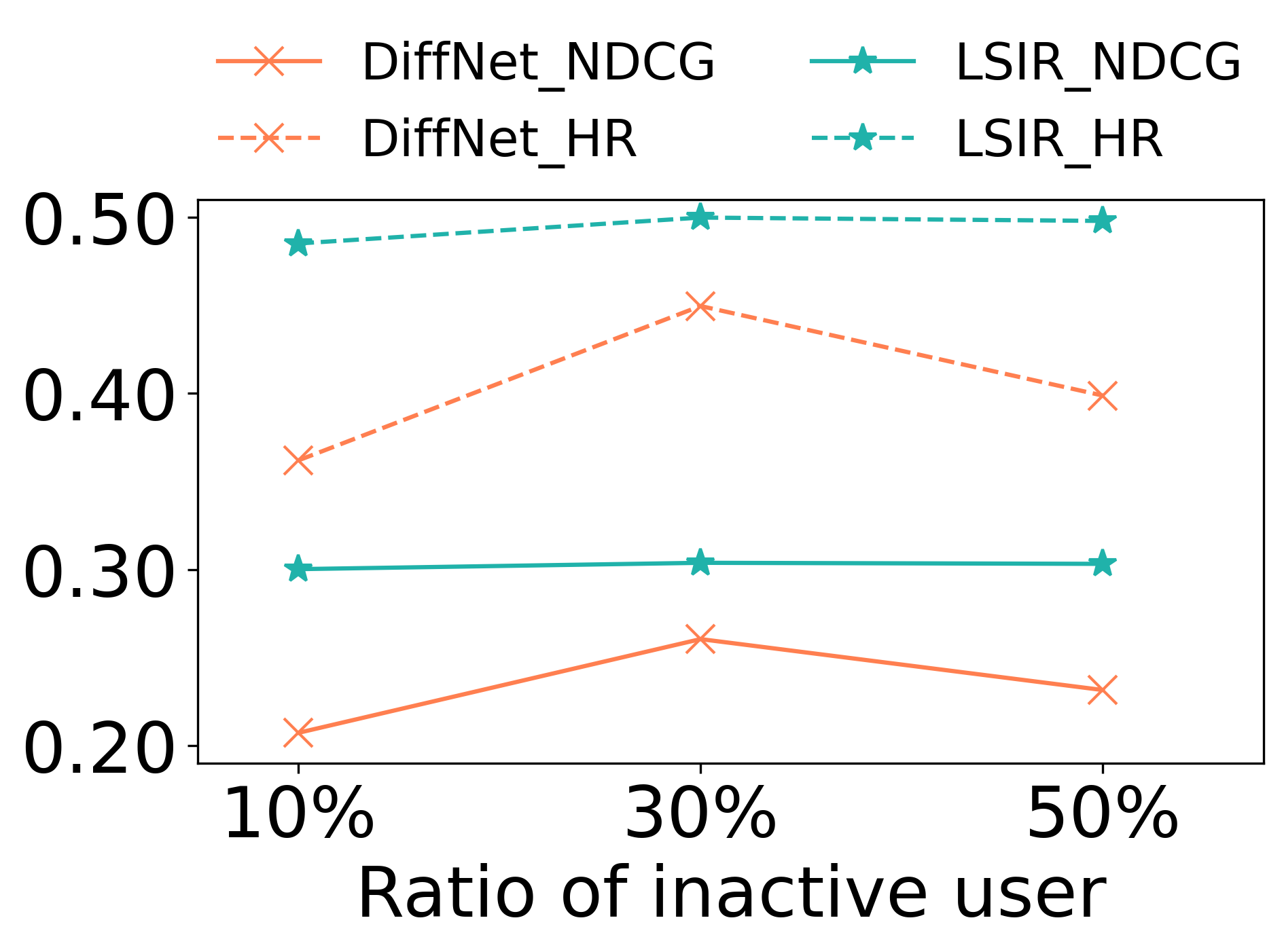}
    \end{minipage}
}
\caption{Test on the ratio of inactive users.}
\label{ratio_inac}
\end{wrapfigure}

Considering there is not definition of `inactive' in selected public datasets (i.e., Flickr and Yelp), we also label the last 30\% users as `inactive' in Table~\ref{exp_stat}. To address any concerns regarding ratio selection, we provide a comparison between DiffNet++ (the best baseline) and LSIR on Yelp, considering different ratios of inactive users. As shown in Fig.~\ref{ratio_inac}, LSIR outperforms DiffNet++ consistently and exhibits greater stability across various ratios of inactive users. Therefore, the selection of ratio will influence less on the performance of LSIR.



\subsubsection{Analysis of Cluster $\mathcal{C}$} 
\begin{wrapfigure}[9]{r}{0.6\textwidth}
\centering
    \subfigure[Flickr]
{
    \begin{minipage}[b]{.25\textwidth}
        \centering
        \includegraphics[width=\textwidth]{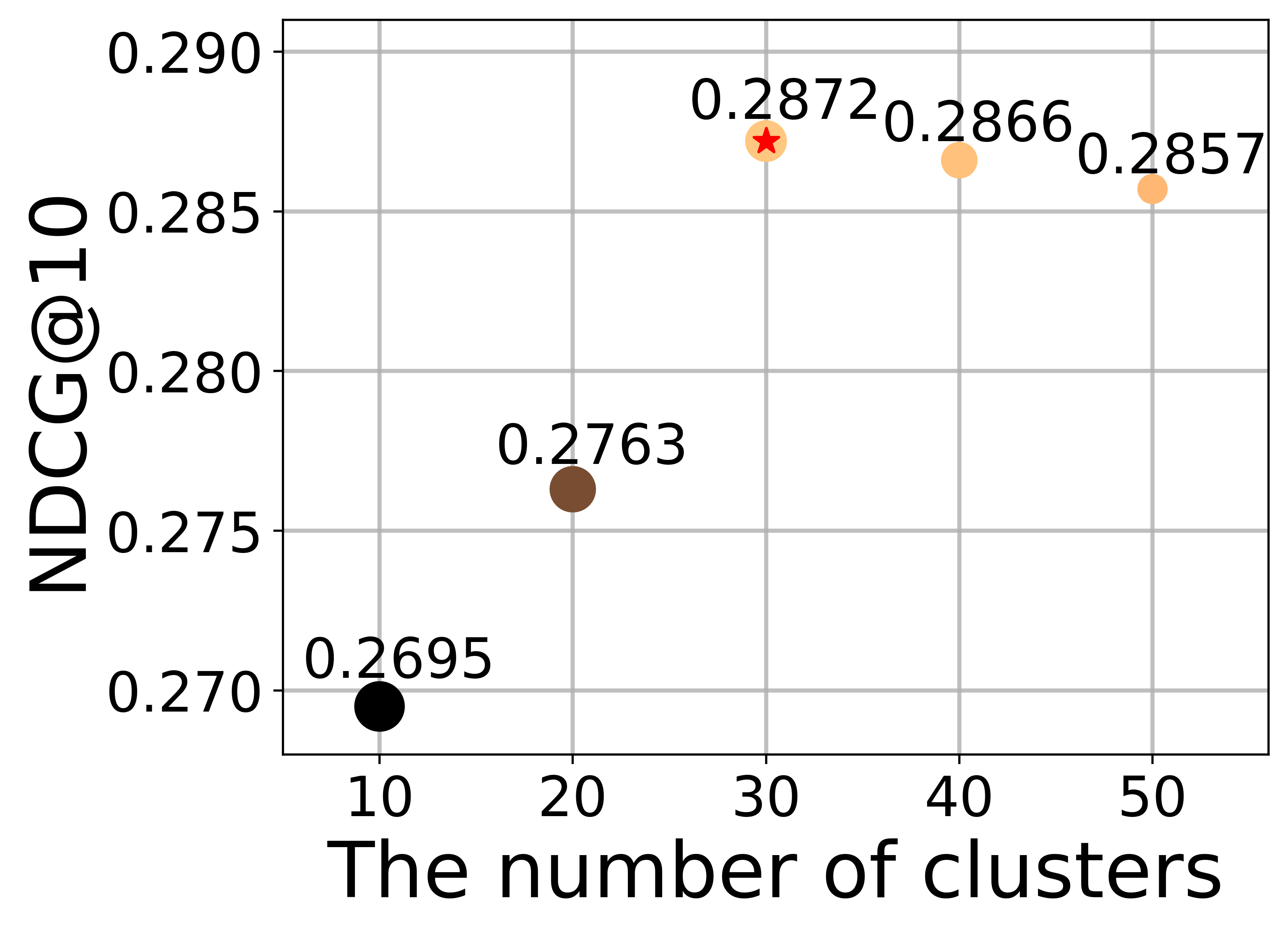}
    \end{minipage}
}
\subfigure[Yelp]
{
 	\begin{minipage}[b]{.25\textwidth}
        \centering
        \includegraphics[width=\textwidth]{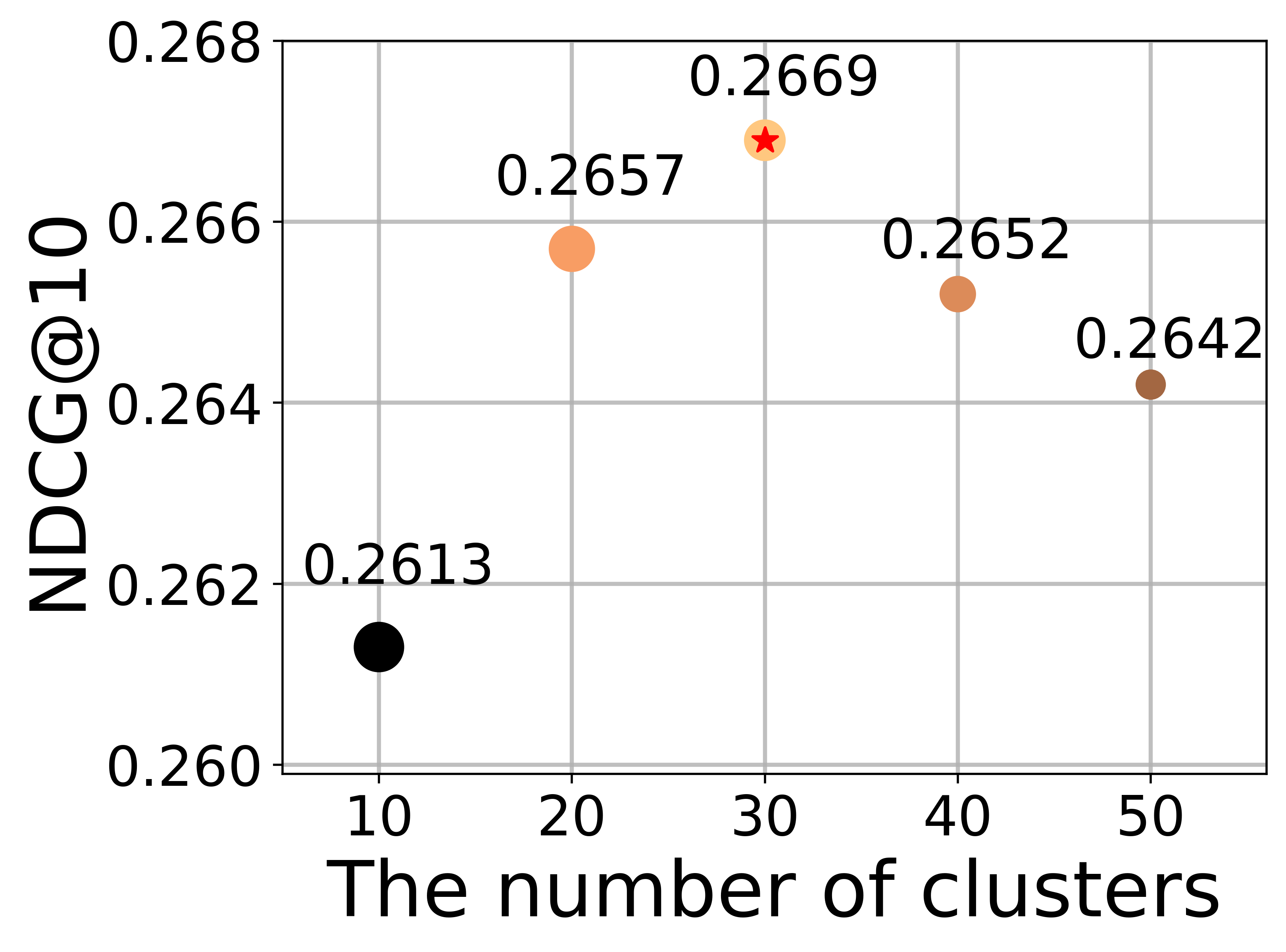}
    \end{minipage}
}
\caption{Test on the number of cluster $\mathcal{C}$.}
\label{c_num}
\end{wrapfigure}

In this analysis, we examine the sensitivity of number of cluster $\mathcal{C}$ utilized in active interest clusters, plotted in Fig.~\ref{c_num}. In this figure, points with lighter colors indicate better performance, while the best point is denoted by a red star. We find that the optimal number of clusters for both the Flickr and Yelp datasets is 30. Using a lower number of clusters would result in an inadequate representation of the diversity of active interests. Conversely, employing a higher number of clusters would make it more challenging for inactive users to find a suitable cluster.

\section{Related Work}
\label{Related Work Supplement}
\textbf{Social recommendation}. GNN based social recommenders~\cite{sharma2022survey} has emerged as a promising direction in recent years. GraphRec~\cite{graphrec} directly utilizes attention strategy to aggregate neighbors for both users and items. DiffNet++~\cite{diffnet++} captures the recursive dynamic social diffusion in users and items. DESIGN~\cite{DBLP:conf/www/TaoLZHW22} jointly models user-item and user-user relations, and enhances it with distillation from separately modeling these two relations. GDSRec~\cite{DBLP:journals/tkde/ChenXLHL23} designs a decentralized interaction graph to vectorize biases from users and items. DH-HGCN~\cite{DBLP:conf/sigir/HanTTX22} constructs hypergraphs separately for each social relation and conducts hypergraph convolution on item-item similar graphs. $S^2$-MHCN~\cite{mhcn} applys multi-channel hypergraph to enhancing mutual information. DcRec~\cite{DBLP:conf/cikm/WuFCL0T22} disentangles user behaviors into social domain and collaborative domain, and utilizes contrastive learning to transfer knowledge from the former to the latter. SDCRec~\cite{DBLP:conf/sigir/DuY00022} utilizes contrastive learning to eliminate popularity bias in cold-start recommendation. SEPT~\cite{sept} combining tri-training strategy and self-supervised learning. The above methods only operate on the raw social graph without any refinement. Inversely, our LSIR carefully modifies social graph to mitigate the mentioned flaws, and specifically benefits inactive users.

\textbf{Cold-start Recommendation}. Cold-start recommenders~\cite{gope2017survey} mainly rely on meta learning. For example, MetaHIN~\cite{metahin} incorporates heterogeneous information to capture rich semantics of meta-paths. MAMO~\cite{mamo} introduces two feature-specific and task-specific memory matrix based on user profiles. MeLU~\cite{melu} deploys the framework of MAML~\cite{maml} in recommendation. TaNP~\cite{tanp} utilizes meta-learning, designing a task-adaptive neural process to enhance cold-start user modeling.  Nevertheless, these meta learning based models are not rely on social connections, which immensely limits their performance.

\textbf{Graph Structure Learning}. Graph structure learning aims to estimate a better structure for original graph~\cite{zhu2021survey}. LDS \cite{lds} jointly optimizes the probability for each node pair and GNN in a bilevel way. Pro-GNN \cite{prognn} aims to obtain a clear graph by deploying some regularizers, such as low-rank, sparsity and feature smoothness. SUBLIME~\cite{sublime} explores the possibility to learn a good structure in unsupervised settings with the help of graph contrastive learning. Unfortunately, these models only focus on node classification task, while little effort has been made towards improving social recommendation.

\setcounter{footnote}{0}
\section{Conclusion}
In this paper, we systematically observe that real social relations are both insufficient quantity and inferior quality for inactive users. 
We take an initial attempt to incorporate the graph structure learning into refining social graph for inactive users. 
By deleting noisy connections and establishing useful links towards active interest clusters of social users, our model LSIR can well alleviate the cold start problem for inactive users, showing the superiority of learning from social structure in recommender systems.

\begin{credits}
\subsubsection{\ackname} This work was supported by Alibaba Group through Alibaba Innovative Research Program.

\subsubsection{\discintname}
The authors have no competing interests to declare that are relevant to the content of this article. 
\end{credits}

\bibliographystyle{splncs04}
\bibliography{sample-base}

\end{document}